\documentclass[fleqn,usenatbib]{mnras}

\usepackage[T1]{fontenc}

\usepackage{graphicx,amsmath,amssymb,makecell}
\usepackage{newtxtext,newtxmath}

\DeclareRobustCommand{\VAN}[3]{#2}
\let\VANthebibliography\thebibliography
\def\thebibliography{\DeclareRobustCommand{\VAN}[3]{##3}\VANthebibliography}

\def\thebibliography{\DeclareRobustCommand{\VAN}[3]{##3}\VANthebibliography}

\title[MOOSE I: Initial Spectral Results]{Monitoring Observations of SMC X-1's Excursions (MOOSE) I: Program Description and Initial High-State Spectral Results}

\author[K. C. Dage et al.]{
Kristen C. Dage$^{1,2}$\thanks{E-mail: kristen.dage@mcgill.ca}
McKinley Brumback,$^{3}$
Joey Neilsen, $^{4}$
Chin-Ping Hu,$^{5}$
\newauthor
Diego Altamirano,$^{6}$
Arash Bahramian,$^{7}$
Philip A. Charles $^{6}$
William I. Clarkson,$^{8}$
\newauthor
Daryl Haggard,$^{1,2}$
Ryan C. Hickox,$^{9}$
Jamie Kennea$^{10}$
\\
$^{1}$Department of Physics, McGill University, 3600 University Street, Montr\'eal, QC H3A 2T8, Canada\\
$^{2}$McGill Space Institute, McGill University, 3550 University Street, Montr\'eal, QC H3A 2A7, Canada\\
$^{3}$ Cahill Center for Astronomy and Astrophysics, California Institute of Technology, 1216 E California Blvd, Pasadena, CA 91125, USA\\
$^{4}$ Villanova University, Department of Physics, Villanova, PA 19085, USA;\\
$^{5}$ Department of Physics, National Changhua University of Education, Changhua, 50007, Taiwan \\
$^{6}$Physics \& Astronomy, University of Southampton, Southampton, Hampshire SO17 1BJ, UK\\
$^{7}$International Centre for Radio Astronomy Research $-$ Curtin University, GPO Box U1987, Perth, WA 6845, Australia\\
$^{8}$Department of Natural Sciences, University of Michigan-Dearborn, 4901 Evergreen Rd. Dearborn, MI, 48128\\
$^{9}$ Department of Physics \& Astronomy, Dartmouth College, 6127 Wilder Laboratory, Hanover, NH 03755, USA\\
$^{10}$ Department of Astronomy and Astrophysics, The Pennsylvania State University, University Park, PA 16802, USA}

\date{Accepted XXX. Received YYY; in original form ZZZ}

\pubyear{2022}

\begin{document}
\label{firstpage}
\pagerange{\pageref{firstpage}--\pageref{lastpage}}
\maketitle

\begin{abstract}
SMC X-1 has exhibited three super-orbital period excursions since the onset of X-ray monitoring beginning with RXTE’s launch in 1995. NICER has recently probed a fourth observed excursion beginning in 2021 with our program Monitoring Observations of SMC X-1’s Excursions (MOOSE). These sensitive new MOOSE data probe different super-orbital periods and phases within them. Spectral fits to the high-state continuum during April 2021 to January 2022 show that the intrinsic spectral shapes are characterised by a soft (kT$\sim$0.19 keV) disc component and a hard ($\Gamma$ $\sim$0.7) power-law tail. When the 2021-2022 NICER observations, taken during an excursion, are compared to 2016 XMM-Newton observations (outside of an excursion), we find little evidence for intrinsic spectral variability across the high-states, but find evidence for a $>$3 $\sigma$ change in the absorption, although we caution that there may be calibration differences between the two instruments. Thus, over different lengths of super-orbital periods, we see little evidence for intrinsic spectral changes in the high-state.  Upcoming studies of the pulse profiles may shed light on the mechanism behind the excursions.

\end{abstract}

\begin{keywords}
accretion, accretion discs – stars: pulsars: individual: SMC X-1 – X-rays:
binaries\end{keywords}



\section{Introduction}
Modulation in X-ray binary lightcurves has long been used as a diagnostic to understand key physical processes in the system. One such modulation that has garnered interest over the last $\sim$50 years are observations of super-orbital periodicities.

Super-orbital periods$--$ the detection of a periodicity in an X-ray binary which is longer than the orbital period of the binary$-$ have been significantly detected in many different types of systems. These include supergiant fast X-ray transients, wind-accretion systems, as well as persistent (Roche lobe overflow), high-mass X-ray binaries and low-mass X-ray binaries with both black hole and neutron star primaries \citep{Kotze12, Corbet13}. Recent optical studies also showed evidence for super-orbital periodicites in the black hole transient MAXI J1820+070 \citep{Thomas2022}.

One explanation is that super-orbital period modulation is due to a warped, precessing accretion disc that is observed edge-on, such that the disc obscures the accretor during precession \cite{Clarkson03b}. As such, the super-orbital period is a direct observable of the accretion disc configuration. For systems like the neutron star X-ray binary SMC X-1, one generally accepted cause of warps in X-ray binary accretion discs is radiation driven warping \citep{Pringle1996, Ogilvie2001}. The stability of the warped disc structure is determined by binary parameters such as radius and mass ratio \citep[e.g.][]{Clarkson03, Clarkson03b}. In systems that do not fall in the stability range, like SMC X-1, fluctuations in the geometry of the accretion disc can be observed as fluctuations in super-orbital behaviour \citep{Ogilvie2001}.

SMC X-1 is a well-studied high mass X-ray binary with a known neutron star accretor and a B0 I supergiant companion \citep{Reynolds93}, The system has a 3.89 day orbital period \citep{Schreier72} and the neutron star produces a $\sim$ 0.7s pulse period \citep{luck76}, which, since the 1970s, its $\sim$ 0.7 s pulse period has steadily been spun up \citep{Inam2010}. \cite{Hu19} report the spin-up frequency derivative as $\sim$ 2.5 $\times 10^{-11}$ s$^{-2}$. Early studies of SMC X-1's X-ray spectral properties identified both hard and soft spectral components \citep{marshall83}. These components likely originate from different physical processes \citep{Naik04}. \cite{Hickox05} modeled the hard and soft emission with a twisted inner disc, which indicated that the hard emission may be due to the illuminating pulsar beam, and the soft component caused by the reprocessed emission of the inner accretion disc. 

SMC X-1 is also well known for its super-orbital period, which is typically measured close to 60 days, but can take excursions to periods as low as 45 days \citep{Clarkson03,Dage19, Hu19}. The characterisation and physical origins of the super-orbital period ($P_{sup}$) modulation have been the subject of many studies over the last few decades \citep[see, e.g.][and references therein]{Gruber84, Wojdowski98, Clarkson03, Trowbridge07, Hu2011,Hu2013, Dage19, Hu19}, with many techniques implemented to accurately measure these long-term modulations, including dynamical power spectrum and Hilbert-Huang \citep{Huang98} transform analysis.

Spectral studies by \cite{Wojdowski2000} determined that the spectral properties are at odds with predictions from line-driven wind models \citep{Blondin95}. RXTE spectral monitoring by \cite{Inam2010} found little evidence of spectral variation, except that $N_\textrm{H}$ increased as the X-ray flux dropped, and monitoring of the spectrum by \cite{Vrtilek05} showed that the super-orbital period cycle influences the spectrum, as the low state spectra were linked to the orbital phase, but not the high state spectra. The role that radiation-driven warping plays in SMC X-1's super-orbital period is very much a mystery, as recent analysis by \cite{Pradhan20} suggests that a warped accretion disk is not the only cause of $P_{sup}$, but variability in the power-law normalization could imply that intrinsic changes from the source itself are a factor in producing the observed super-orbital periodicity.

Other tests to determine the geometry of the disc can also be probed via studies of the pulse profiles: variation in the soft pulse profiles may be related to precession in the disc \citep{Neilsen04}. \citet{Brumback20} implemented X-ray tomography techniques to directly model the warped accretion disc, in particular, the inner part near the magnetosphere, and showed that the direct (hard) pulse profile and reprocessed (soft) pulse profile can both be separately traced by X-ray spectra. The hard and soft pulse profile offsets changed within a given 60 day super-orbital period cycle, but the pulse profiles at the beginning and end of the 60 day cycle were almost identical$-$ implying that the inner accretion disc had completed an entire precessed cycle. 

Observational and theoretical tests can be made to determine if SMC X-1's disc warp and roving super-orbital period is attributed to radiation driven warping, where irradiation from a central source warps the accretion disc. Smoothed particle hydrodynamical simulations by \cite{Foulkes06, Foulkes2010} suggest this is quite possible. Such a scenario would imply a link between the \textit{change} in super-orbital period and the mass transfer rate. \cite{Dage19} searched for such a link, but they were not able to find conclusive observable evidence. \cite{Hu19}'s complete analysis of the super-orbital period from 1999-2019, along with analysis of spin periods thanks to \textit{MAXI GSC} found no evidence that the mass transfer rate was strongly associated with the super-orbital period, but suggested that the super-orbital period modulation could be due to a torque change (which in turn is dependent on the mass transfer rate and the warp angle).

Understanding the nature of SMC X-1's warped, precessing accretion disc has risen to greater importance with the discovery of the pulsating ultraluminous X-ray sources (ULXPs; e.g., \cite{Bachetti14}). SMC X-1 may be a nearby analog to the ULXPs, as it also exhibits pulse dropouts \citep{Pike19, Bachetti2020}. Many ULXs are now thought to  exhibit super-orbital periods and possibly also accretion disc warping \citep[see, e.g.,][among others]{Motch14, Brightman19, Walton16}, see also \cite{Townsend20} and references therein. Understanding SMC X-1's super-orbital behaviour can shed light on the mysterious, and more distant ULXPs, as well as provide a unique understanding on the physical processes responsible for X-ray binary variability on many timescales.

Our \textit{NICER} observing program, \textbf{M}onitoring \textbf{O}bservations \textbf{O}f \textbf{S}MC X-1's \textbf{E}xcursions (MOOSE) is designed to fully utilise \textit{NICER}'s monitoring capabilities to provide information on SMC X-1's spectral shape and soft timing properties as the sources enters a new state of super-orbital period modulation.  These properties have not yet been probed during an epoch of super-orbital period excursion, and MOOSE will provide a rich and diverse data-set to shed light on the physical properties of the system during this time. This paper is the first in the series focusing on the high-state spectra. Detailed timing analysis of the system will be presented in Brumback et al. (in prep). In Section \ref{section:monitoring} we describe the MOOSE campaign; Section \ref{section:data} discusses initial spectral analysis. These results are discussed in Section \ref{section:results}, while Section \ref{section:discussion} considers the next steps to be undertaken in the analysis of MOOSE's dataset.

\section{Data and Analysis}
\label{section:data}
\begin{figure*}
    \centering
    \includegraphics[width=6.5in]{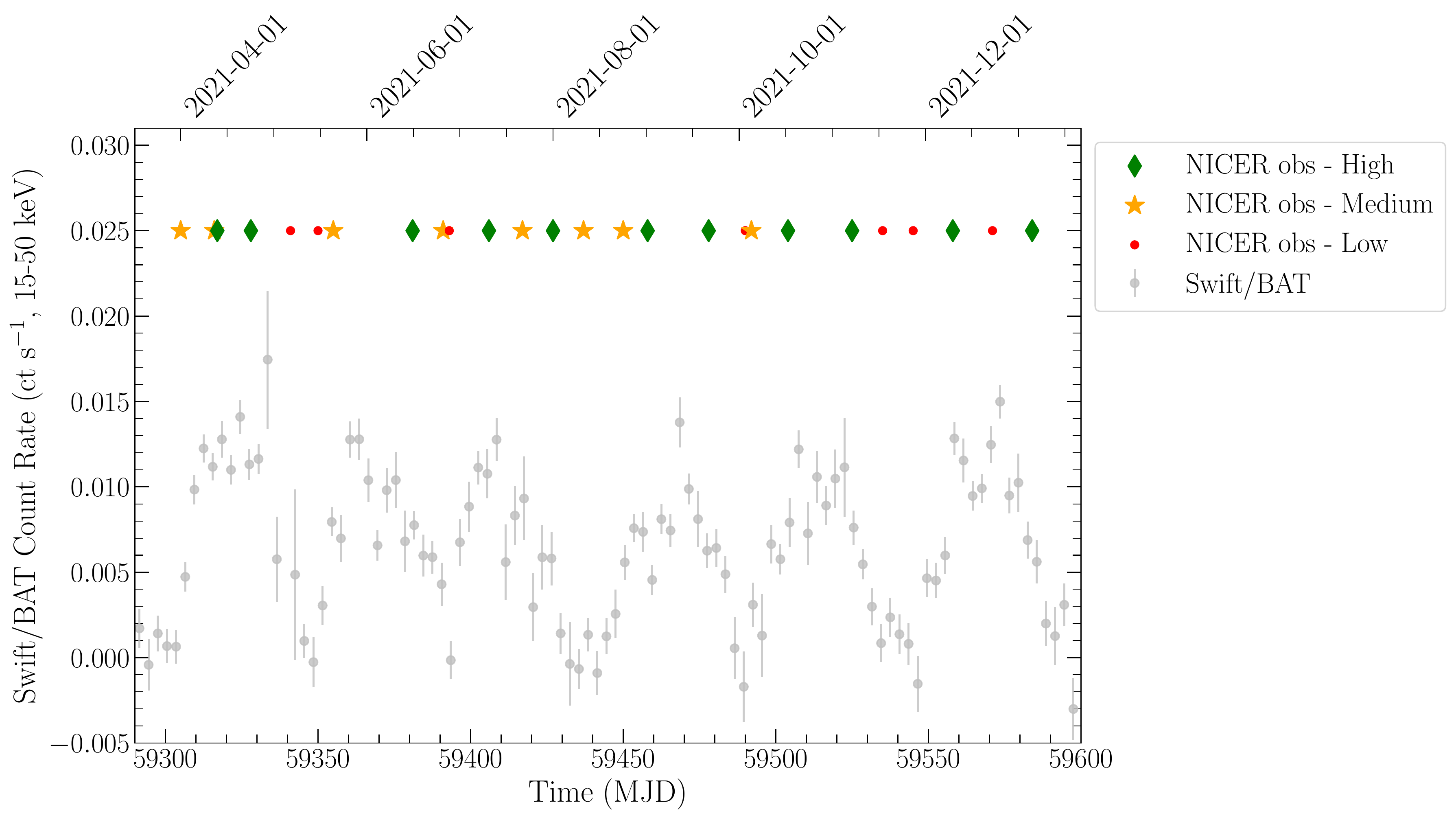}
    \caption{Rebinned \textit{Swift}/BAT monitoring of SMC X-1 overlaid with observation times from the \textit{NICER} monitoring campaign. Each observation is differentiated by data quality (green diamond are the highest count rates, yellow star is intermediate, red dot is the lowest), which is a function of observation length and where within the super-orbital period cycle the snapshot was taken.}
    \label{fig:observations}
\end{figure*}

\textit{NICER}'s superb timing resolution ($<$300 nsec) and monitoring capabilities, along with sensitivity in the soft (0.2-12 keV) energy range \citep{Gendreau} present an excellent opportunity to study SMC X-1 throughout the duration of the newest super-orbital period excursions. The observations are spaced roughly 10 days apart, and sample different stages of the super-orbital cycle. The data quality is a function of the observation length and the phase of the super-orbital cycle that is observed (as observations taken in the low state have a much lower count rate). Figure \ref{fig:observations} shows the observations plotted against the \textit{Swift}/BAT monitoring lightcurve, with each observation flagged by data quality. 

As discussed in Section \ref{subsection}, the super-orbital cycles are not consistent with each other for the duration of the monitoring interval (see also Figure \ref{fig:observationsnicer}).
\subsection{Data Reduction and Reprocessing}
\label{section:monitoring}

The \textit{NICER} observations were processed with HEAsoft 6.29 \textsc{nicerl2} task, with the overonly range extended to 1.5, as the overonly ranges of many of the data often exceeded this value (see Table \ref{table:obs}). Overshoots (the overonly range) measure the high energy particle backgrounds, while undershoots (the underonly range) are detector resets which measure red noise and any optical light. \footnote{\url{https://heasarc.gsfc.nasa.gov/docs/nicer/data_analysis/workshops/NICER-Workshop-Filtering-Markwardt-2021.pdf}} The spectra were extracted with \textsc{xselect}, and we implemented optimal binning \footnote{\url{https://heasarc.gsfc.nasa.gov/docs/nicer/analysis_threads/spectrum-grouping/
}} with no minimum number of counts \citep{kaastrebleeker}. The background spectra were generated using the \textsc{nibackgen3C50} tool \citep{Rem21}, and we used the \textsc{nicerarf} and \textsc{nicerrmf} tasks to generate individual ARF and RMF files for each spectrum.

\begin{table*}
\caption{\textit{NICER} observations log.}
\label{table:obs}
\begin{tabular}{rrrrrr}
\hline
ObsID      & Date       & Duration  & Overonly Range & Underonly Range & Count Rate  \\ 
&&(sec)&&(ct/sec)\\ \hline
4509012401 & 2022-01-05 & 1474 & \textless{}15 &  \textless{}300& 151\\
4509012301 & 2021-12-23 & 2411         & \textless{}1   & \textless{}306  &      5               \\
4509012201 & 2021-12-10 & 1476         & $<$11           & \textless{}200  & 249                 \\
4509012101 & 2021-11-27 & 1513         & $<$100          & \textless{}200  & 6           \\
4509012001 & 2021-11-17 & 955          & $<$75           & \textless{}200  & 10                 \\
4509011901 & 2021-11-07 & 1259         & $<$30           & \textless{}200  & 176                 \\
4509011701 & 2021-10-17 & 2533         & \textless{}1   & \textless{}200  & 195                 \\
4509011601 & 2021-10-05 & 1107         & $<$10           & \textless{}200  & 10                  \\
4509011501 & 2021-10-03 & 350          & $<$16           & \textless{}200  & 9                 \\
4509011401 & 2021-09-21 & 1476         & $<$35           & \textless{}200  & 245                 \\
4509011301 & 2021-09-01 & 1898         & $<$0.5          & \textless{}200  & 189                 \\
4509011201 & 2021-08-24 & 1327         & $<$20           & \textless{}200  & 52                  \\
4509011101 & 2021-08-11 & 1123         & $<$1            & \textless{}200  & 9                 \\
4509011001 & 2021-08-01 & 1235         & $<$50           & \textless{}200  & 72                  \\
4509010901 & 2021-07-22 & 821          & $<$20           & \textless{}200  & 179 \\
4509010803 & 2021-07-11 & 944  & $<$30  & \textless{}200 & 232  \\
4509010802 & 2021-06-28 & 796  & $<$20  & \textless{}200 & 10   \\
4509010801 & 2021-06-26 & 1684 & $<$1.5 & \textless{}200 & 139  \\
4509010701 & 2021-06-16 & 1279 & $<$2   & \textless{}200 & 206  \\
4509010601 & 2021-05-21 & 1024 & $<$4   & \textless{}200 & 146  \\
4509010501 & 2021-05-16 & 1446 & $<$140 & \textless{}200 & 7  \\
4509010401 & 2021-05-07 & 967  & $<$40  & \textless{}200 & 2 \\
4509010301 & 2021-04-24 & 436  & $<$20  & \textless{}200 & 195  \\
4509010103 & 2021-04-13 & 483  & $<$7   & \textless{}200 & 220  \\
4509010102 & 2021-04-12 & 1362 & $<$3   & \textless{}200 & 214  \\
4509010101 & 2020-04-01 & 1006 & $<$11  & \textless{}200 & 33\\ \hline
\end{tabular}
\end{table*}

\subsection{Determining changes in $P_{sup}$ from BAT monitoring}
\label{subsection}
SMC X-1 has been observed to undergo three separate epochs of super-orbital period excursion, which last for 2-3 years, occurring in 1997, 2006, and 2014 \citep{Hu19}. While it is difficult to accurately predict the onset of the next super-orbital period excursion, the $\sim$ 6 year difference gap between previous super-orbital period excursions suggested that SMC X-1 would undergo some type of super-orbital period excursion beginning in 2021. 

Most techniques to perform accurate measurements of super-orbital periodicities, such as the Lomb-Scargle periodogram \citep{Scargle82} often require hundreds of days worth of data to perform a single period measurement. While there is always a concern that red noise can cause problematic false detections \citep{vaughan16}, SMC X-1's super-orbital period is well-known and has been observed with many repetitions. 

Given that our monitoring campaign has not yet exceeded one year and is comprised of short observations with days-long gaps, techniques such as the Lomb-Scargle periodogram are not the most effective methods to constrain the super-orbital period. However, phase dispersion minimization is one of the techniques enabling us to measure periods from these data sets. 

\begin{figure*}
    \centering
    \includegraphics[width=5in]{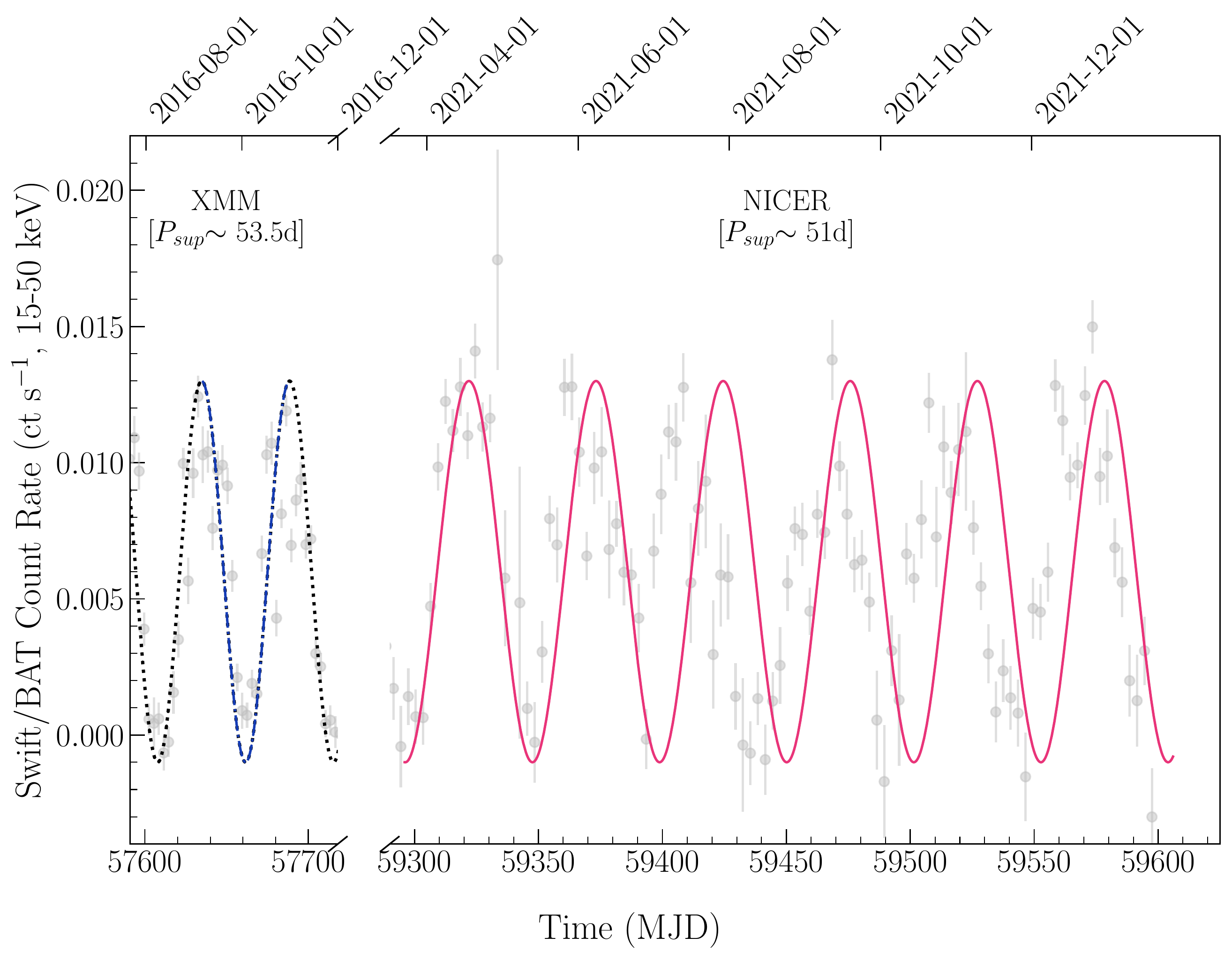}
    \caption{Comparison of the super-orbital period cycle during the 2016 \textit{XMM-Newton} observations and the current \textit{NICER} monitoring campaign. We fit the two different epochs of \textit{Swift}/BAT observations with P4J and overlay the model with best-fit parameters on the image.  The dashed blue lines on the left panel are the cycle that was observed by \textit{XMM-Newton}, while the dotted lines show the best-fit period over the whole data-set used to measure the period. The left panel shows the P4J model during the \textit{NICER} monitoring. }
    \label{fig:observationsnicer}
\end{figure*}

To quantify any super-orbital period changes between the 2016 and 2021 epochs, we implement the python package P4J, where the criterion for period detection is based on maximizing the Cauchy-Schwarz Quadratic Mutual Information \citep{huijse18} \footnote{\url{https://github.com/phuijse/P4J}} to measure the best super-orbital period of the BAT data taken from MJD 59290-59606, which covers 5 super-orbital cycles (we removed the cycle before MJD 59290 because the data gap was problematic), and interpolated a sine model from the best fit period ($P_{sup}$=51d). For comparison, we implemented the same methodology on 5 cycles of the BAT data from MJD 57535-57788 (during the time the comparison \textit{XMM-Newton} observations were taken). P4J returned the best fit period of 53 days, and the 2016 cycles appear to be self-consistent.  We superimpose these models over-top the \textit{Swift}/BAT monitoring in Figure \ref{fig:observationsnicer}. While the \textit{XMM-Newton} epoch can be represented by a sinusoidal model with the period derived from P4J, the \textit{NICER} epoch cannot all be fit by the modeled P4J period for that epoch, which implies that the super-orbital period is not a constant value during this time. This suggests that the MOOSE data set is unique because it provides both spectral and timing windows into unprobed regimes of $P_{sup}$ for SMC X-1. 

\subsection{Monitoring of the High-State Continuum during the excursion}
\begin{figure*}
    \centering
    \includegraphics[width=5.5in]{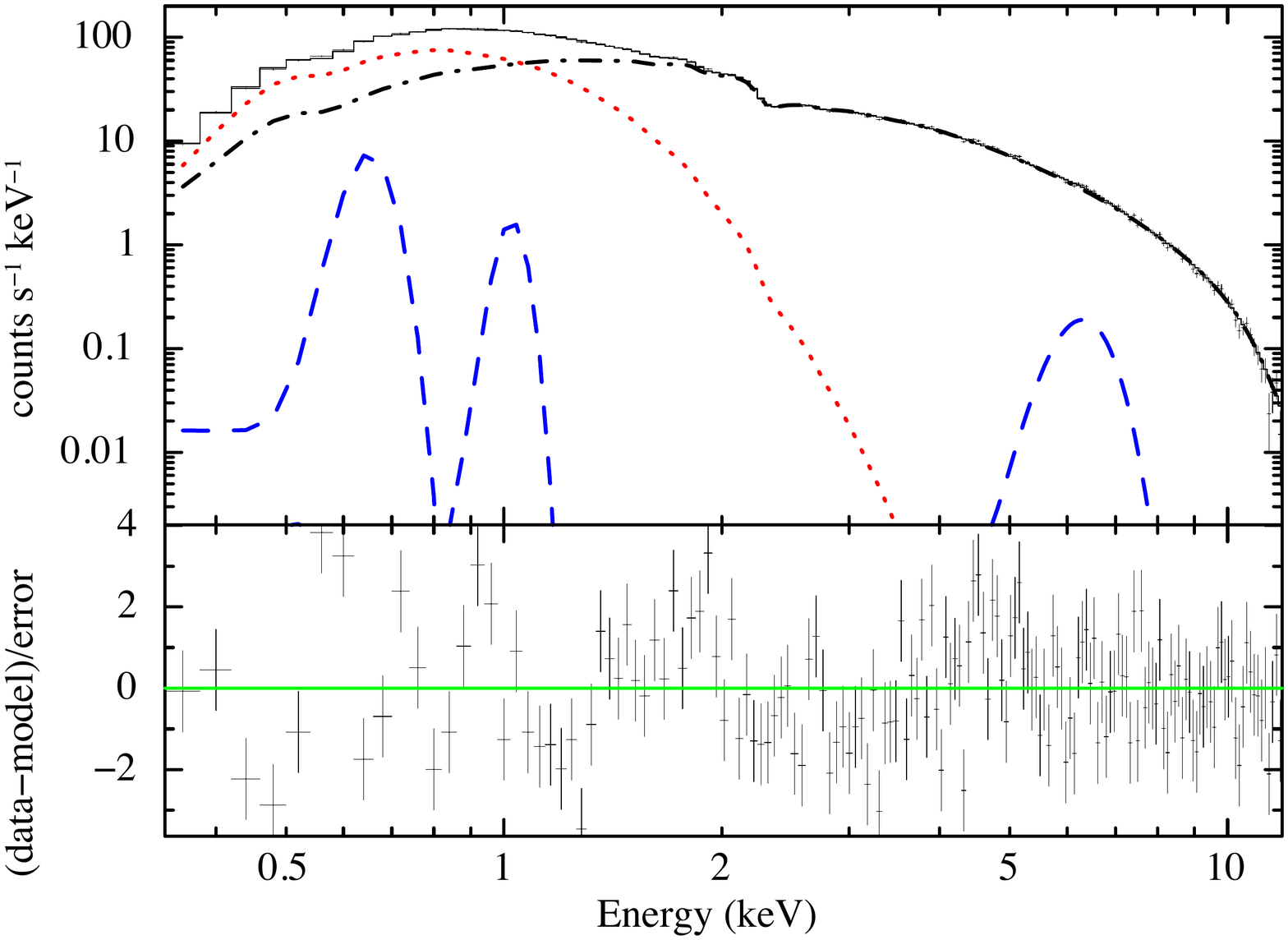}
    \caption{Best-fit spectrum and residuals of ObsID 4509011701. The spectrum was fit with tbabs*(bbody+cutoffpl+gaussian+gaussian+gaussian). The Gaussian components for line energy and  $\sigma$ were fixed to the values from \citet{Brumback20}. The \textsc{xspec} model components are as follows: bbody (red dotted line), cutoffpl (black dash-dotted line), and the three Gaussian components (blue dashed line). The uncertainties in the \textit{XMM-Newton} observations are likely underestimated given the high reduced $\chi^2$ values.}
    \label{fig:spec}
\end{figure*}

\begin{figure*}
    \centering
    \includegraphics[width=5in]{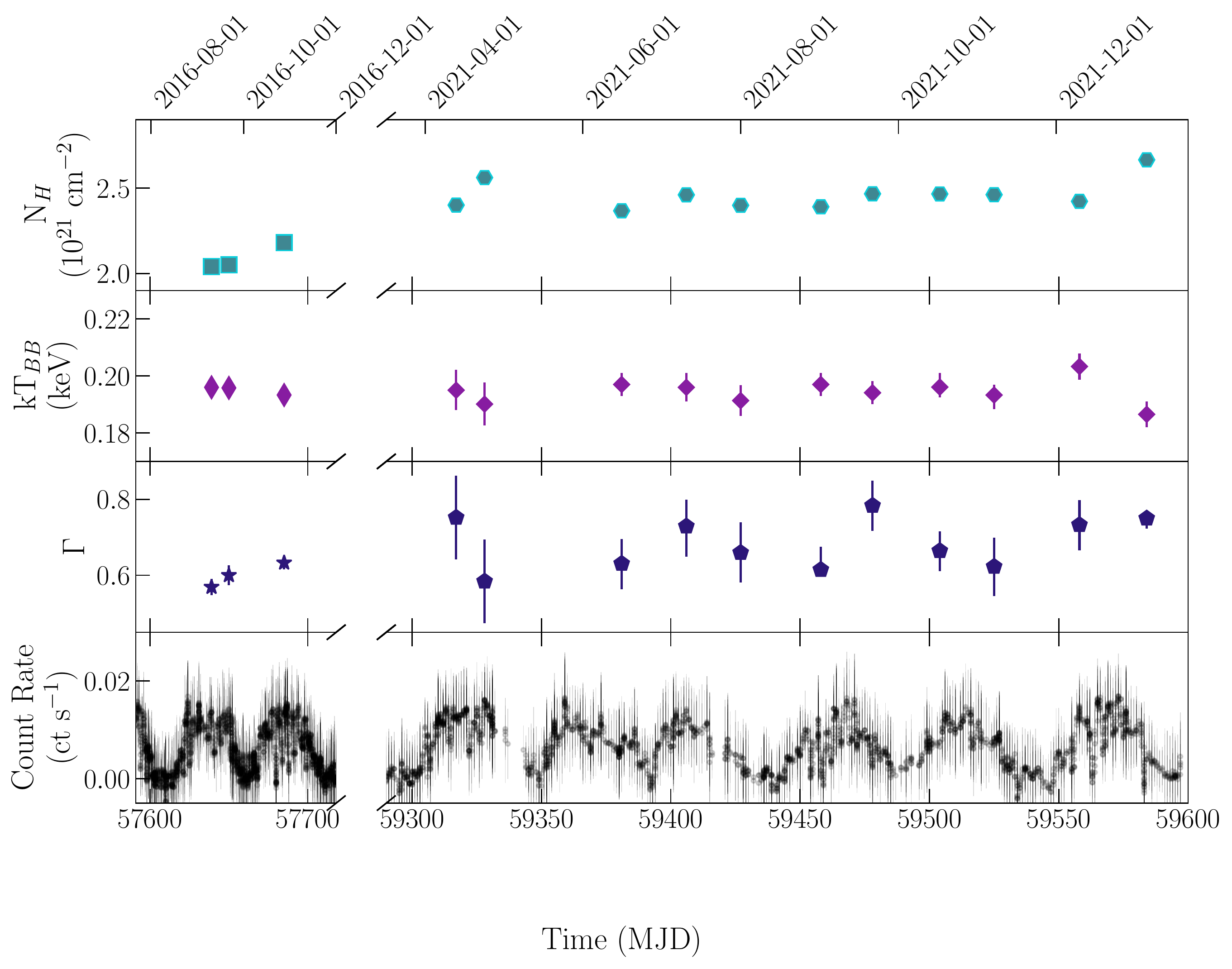}
    \caption{Best-fit parameters of $N_\textrm{H}$, kT and $\Gamma$ for high-state observations, with \textit{Swift}/BAT light curve for reference. These parameters can be found in Table \ref{table:fits}. }
    \label{fig:bestfits}
\end{figure*}
As an initial step to monitor the continuum, we fit spectra of the high-state observations using \textsc{xspec} version 12.12.0, with solar abundances from \cite{Wilm} and $\chi^2$ statistics, with photo-electric absorption cross-sections from \cite{Verner1996}. All reported uncertainties are at 90\% confidence, and we implemented optimal binning \footnote{\url{https://heasarc.gsfc.nasa.gov/docs/nicer/analysis_threads/spectrum-grouping/}}. The high-state observations can be described by the model \texttt{tbabs*(bbody+cutoffpl)} with three \texttt{gaussian} components added to model emission lines (see Figure \ref{fig:spec}). We restricted the search range of the high-energy cut-off value between 12 keV and 20keV to allow the fits to more closely match \citet{Brumback20}. While \textit{NICER} does not extend to this range, the inclusion of a hard component is necessary to fit the spectrum. SMC X-1 is known to produce emission from Fe K$\alpha$ (6.4 keV, broad),  Ne X Ly$\alpha$ (1.2 keV), and O VIII Ly$\alpha$ (0.65 keV), and we fix our emission lines to the same line energy and width as \citet{Brumback20}. The low/intermediate state spectra are not well-fit by this model. We will undertake a comprehensive examination of all of the low/intermediate state data once the full MOOSE dataset has been acquired.

For comparison, we fit the same models to the high state \textit{XMM-Newton} observations from \citet{Brumback20} (ObsIDs 0784570201, 0784570301, 0784570501; see \citealt{Brumback20} for analysis details), which were obtained in 2016 when the source was not in a state of super-orbital period modulation.

The best-fit parameters for $N_\textrm{H}$, kT and $\Gamma$ are presented in Table \ref{table:fits} and can be seen in Figure \ref{fig:bestfits}. We note that given the spectral resolution of \textit{XMM-Newton}, the models we use are likely over-simplistic, however, we use the same models for \textit{XMM-Newton} and \textit{NICER} for sake of consistency and ease of comparison across the epochs. 
\begin{table*}
\caption{Best-fit spectral parameters for high-state observations from each super-orbital period epoch. The \textit{NICER} and \textit{XMM-Newton} observations are fit in the 0.3-12 keV energy range.}
\label{table:fits}
\begin{tabular}{rrrrrr}
\hline
ObsID      & Date       & $N_\textrm{H}$ (10$^{21}$cm$^{-2}$) & kT (keV)          & $\Gamma$   &    $\chi^2$ /d.o.f.\\ \hline
07845770201&2016-09-08&2.02 $\pm$ 0.05 &0.197 $\pm$ 0.002 &0.569 $\pm$ 0.021 & 1043.28/208 \\
07845770301&2016-09-19&2.02 $\pm$ 0.06 &0.198 $\pm$ 0.003 &0.603 $\pm$ 0.026&888.42/206\\
07845770501&2016-10-24& 2.1 $\pm$ 0.04 & 0.195 $\pm$ 0.002 & 0.63 $\pm$ 0.018&2026.07/216\\
4509010103 & 2021-04-13 &2.4 $\pm$ 0.1 & 0.20 $\pm$ 0.01& 0.75 $\pm$0.11 &172.35/138\\ %
4509010301 & 2021-04-24 & 2.6 $\pm$ 0.1 & 0.19 $\pm$ 0.01  & 0.58 0.11      &152.06/134\\ %
4509010701 & 2021-06-16 & 2.4 $\pm$ 0.1 & 0.20 $\pm$ 0.01  & 0.63 $\pm$ 0.07 & 286.20 /153\\ %
4509010803 & 2021-07-11 & 2.5 $\pm$ 0.1 &0.20 $\pm$ 0.01& 0.73 $\pm$ 0.08 &191.66/150\\%
4509010901 & 2021-08-01 & 2.4 $\pm$ 0.1  & 0.19 $\pm$ 0.01  & 0.66 $\pm$ 0.08  &198.70/147\\%
4509011301 & 2021-09-01 &2.4 $\pm$ 0.1 & 0.20 $\pm 0.01$ & 0.61 $\pm$0.01 &211.42 /153\\%
4509011401 & 2021-09-21 & 2.5 $\pm$ 0.1  & 0.19 $\pm$ 0.01  & 0.78 $\pm$ 0.07 &272.90/152\\%
4509011701 & 2021-10-17 & 2.5 $\pm$ 0.1  & 0.20 $\pm$ 0.01  & 0.66 $\pm$0.05 &299.49/151\\%
4509011901 & 2021-11-07 & 2.5 $\pm$ 0.1 & 0.20 $\pm$ 0.01 & 0.67 $\pm$0.05 &196.29/145\\%
4509012201 & 2021-12-10 & 2.4 $\pm$ 0.1  & 0.20 $\pm$ 0.01 & 0.73 $\pm$0.07 &239.19/150\\
4509012401 & 2022-01-05 & 2.7 $\pm$ 0.1 & 0.19 $\pm$ 0.01 & 0.75 $\pm$ 0.03&208.44/138\\

\hline
\end{tabular}
\end{table*}

\section{Results and Discussion}
\label{section:results}

We fit \textit{XMM-Newton} and \textit{NICER} high state spectra of SMC X-1 through different $P_{sup}$ values and phases and find no significant spectral changes, except for an apparent moderate variability in $\Gamma$. As the best-fit value for $\Gamma$ depends on the high-energy
cut-off, a parameter \textit{NICER} is not sensitive to (as the best-fit values are sometimes outside the NICER energy range), we cannot yet conclude  whether it is variable or not. As
seen in Figure \ref{fig:bestfits}, the best-fit $N_\textrm{H}$ increases between the XMM-Newton and \textit{NICER} observations (from an average \textit{XMM-Newton} value of 2.1 (+/- 0.08) $\times$ $10^{21}$ cm$^{-2}$ to an average \textit{NICER} value 2.5 ($\pm$ 0.01)$\times$ $10^{21}$ cm$^{-2}$ ).  However, the temperature remains relatively stable with both \textit{XMM-Newton} and \textit{NICER} giving an average kT of 0.20 keV to a precision of $<$0.04 keV.  

There appears to be slight variability in $\Gamma$, although that may be due to a degeneracy with the high-energy cut-off parameter, which \textit{NICER} is not sensitive to (see next section and Fig \ref{fig:degeneracy}). Thus, this study cannot comment on any potential variability of the power-law index. Our observations show that the overall spectral shape remains remarkably steady across the different superorbital modulations and phases covered here.

\subsection{Comparison to Previous Spectral Studies of SMC X-1}
Some comparisons can be made between spectral fits to the MOOSE high-state observations and SMC X-1 spectral fits from the literature. In particular, \cite{Naik04} observed SMC X-1 with \textit{BeppoSAX} while SMC X-1 was in the onset of the first observed super-orbital period excursion (early 1997). Their spectra contained similar features to \textit{ASCA}  observations from \cite{Paul02}: a 6.4 keV iron emission line, a power-law component and a soft excess. However, the BeppoSAX observations were not sufficiently long enough to place constraints on variation in the soft excess between \cite{Paul02} and \cite{Naik04}. We find that the MOOSE observations have the same features, and remarkably similar X-ray fits (over the same energy ranges) across several years and different instruments, e.g. \cite{Naik04}, yielding soft component kT values between 0.16-0.19 keV. Fits by \cite{Brumback20} find that within different super-orbital phases, the overall spectral shape of the high-state remains similar, with any changes occurring in the normalization of the Fe line, or small differences in the primary power-law index. 

\begin{figure}
    \centering
    \includegraphics[width=3.3in]{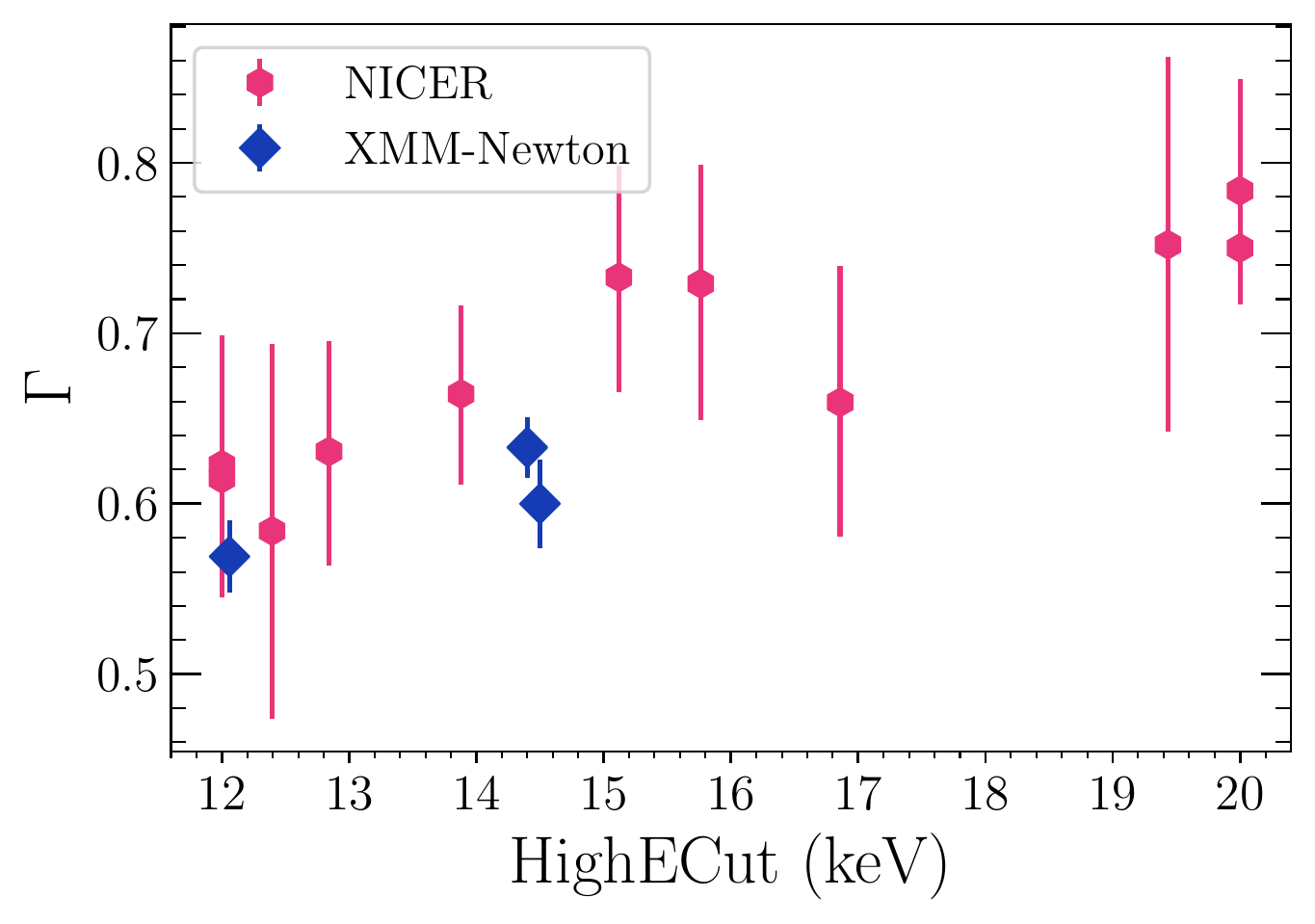}
    \caption{Best fit power-law index compared to best fit high energy cut value. \textit{NICER} is not sensitive to the same energy ranges as \textit{XMM-Newton}, and thus had a wider range of best-fit values for the high energy cut, which in turn impacted the corresponding value for the power-law index.}
    \label{fig:degeneracy}
\end{figure}
The hard bands of SMC X-1 have also been observed many times, with \textit{RXTE}, \textit{Suzaku} and \textit{NuSTAR}: \cite{Inam2010} fit \textit{RXTE}/PCA (3-25 keV) observations with a cutoff power-law only, but find that the only parameter that changes within the error-bars is $N_\textrm{H}$, which is anti-correlated to the X-ray flux. Fits by \cite{Pike19} found that the spectrum in the low-state was poorly constrained due to the low number of source counts, and that a low-temperature (kT$<$ 0.5 keV) component was necessary to model excess soft flux. This component was not recovered by \cite{{Pradhan20}} fits to  \textit{Suzaku} and \textit{NuStar} spectra, but they did find evidence for the hard X-ray spectrum becoming steeper as the flux increased.  

While we do detect some changes in the best fit values for $N_\textrm{H}$ between the 2016 \textit{XMM-Newton} observations and the 2021-2022 \textit{NICER} observations, it is not clear at this point whether the differences are due to calibration (see \citealt{Brumback20} for more detail on the \textit{XMM-Newton} observations), or potential disc changes during excursion. $\Gamma$ also appears to show some variance within the \textit{NICER} observations.However, we note that there is a degeneracy between the best-fit value for the high-energy cut-off, which extends beyond \textit{NICER}'s energy range, and thus it is not clear whether the power-law index shows intrinsic variability, or rather if the fit was not well constrained (see Figure \ref{fig:degeneracy}). Thus, this study is not well-equipped to comment on variability in the power-law index.

\subsection{Comparison of SMC X-1 to ULX Pulsars with Super-Orbital Periods}
SMC X-1 is known to reach high X-ray luminosities that are consistent with being near or above its Eddington limit (for a 1.1 solar mass NS; \cite{bidaud81, Meer07, Pike19}. The richness of observational data on SMC X-1 makes it an ideal target for comparison with extragalactic ultraluminous X-ray pulsars, as several have been found to exhibit evidence for super-orbital periodicities, including NGC 7793 P13; \citep{Motch14}, NGC 5907 ULX1; \citep{Walton16}, M82 X-2; \citep{Brightman19}, M51 ULX7; \citep{Brightman20}. Due to the distance of these galaxies, it is challenging to perform detailed measurements of the pulse profiles, and thus the main observables utilised to understand these systems are the spectral shape and spectral evolution. Thus, it is potentially useful to be able to compare the long-term monitoring of SMC X-1's spectral behaviour to studies of these systems. 

While it is challenging to perform spectral studies of M82 X-2 due to its nearby neighbour \citep{Bachetti2020}, the other three systems have been well-studied. Broad-band spectroscopy by \cite{Walton18} found that the spectrum of NGC 7793 P13 was best fit by two soft components as well as a third, hard continuum component. X-ray spectroscopy of high cadence observations by \cite{Lin22} found that NGC 7793 P13  has both a hard and a soft component, although it is unclear if the source's spectrum shows statistically significant changes over the observations.

\cite{RC20} determined that M51 ULX7 was best fit by a black-body disc with a second (hotter) black-body model. A study of the source's spectral evolution by \cite{Brightman22} found that there is not sufficient evidence for variation of the spectral shape either as a function of super-orbital phase or X-ray luminosity.

Studies of NGC 5907 ULX1 found that the source's spectrum was well characterised by a soft disc component with a power-law tail. Unlike the previous studies of the aforementioned ULXPs, spectral fits to observations of NGC 5907 ULX1 in 2003, 2012, and 2013 suggest that there is some evidence for spectral change as a function of phase \citep{Fuerst17}.

In the new MOOSE spectral monitoring of SMC X-1's high-state continuum, where we have probed a number of different super-orbital phases, we currently do not find strong evidence of intrinsic changes to SMC X-1's high-state continuum spectrum,  although there may be some evidence for changes in $N_\textrm{H}$ or $\Gamma$. Given that past studies have found that many systems find state changes linked to the mass transfer rate \citep[and references therein]{Kotze12}, the lack of clear state changes during SMC X-1's excursion suggest that the relationship between the super-orbital period excursion and the high-state spectral parameters is somewhat complicated.

\section{Summary and Future Studies}
\label{section:discussion}
We present the first full monitoring coverage which affords both high quality spectra and pulse profiles of SMC X-1 as it enters its fourth observed epoch of super-orbital period excursion.  

Our initial findings are summarized as follows:

\begin{itemize}
    \item We find that the new observations probe many different super-orbital phases and different super-orbital period values, and as such will provide a rich and diverse data-set that will be key to understanding the driving causes behind super-orbital period modulations.
    \item Fits to the spectral continuum are best-characterised by an absorbed black body disk with kT $\sim 0.19$ keV, and a hard power-law component ($\Gamma \sim$ 0.7). Despite having probed different super-orbital periods and super-orbital period phases, we do not detect clear and significant evidence of changes to the continuum.
    \item We note a $>$3 $\sigma$ increase in $N_\textrm{H}$ (with kT relatively fixed) from the 2016 XMM-Newton observations to the 2021-2022 \textit{NICER} monitoring, although this could be due to differences in the instruments \citep[e.g.][]{Schellenberger}, rather than any physical changes.  
    \item While the time averaged X-ray spectroscopy is a useful tool to make comparisons with other systems which may have similar accretion mechanisms, we see little evidence for spectral changes to the high-state as a function of super-orbital excursion. 
    \item \textit{NICER} monitoring of SMC X-1 via MOOSE will continue through early 2023. We will release all data products at the end of the monitoring period to cement the legacy value of this program.
\end{itemize}

The next step will be to perform a  full study of pulse profiles (Brumback et al in prep), as well as phase-resolved spectroscopy and detailed study of any variability in the emission lines as a function of orbital and super orbital period, to delve into the physical origins of the super-orbital period modulation.

\section*{Acknowledgements}
The authors thank the anonymous referee for helpful feedback. KCD and DH acknowledge funding from the Natural Sciences and Engineering Research Council of Canada (NSERC), the Canada Research Chairs (CRC) program, and the McGill Bob Wares Science Innovation Prospectors Fund. KCD acknowledges fellowship funding from the McGill Space Institute. DA acknowledges support from the Royal Society. The authors thank Elizabeth Ferrara and the \textit{NICER} team for their efforts with the observation scheduling.
This research has made use of data and/or software provided by the High Energy Astrophysics Science Archive Research Center (HEASARC), which is a service of the Astrophysics Science Division at NASA/GSFC and
NASA’s Astrophysics Data System Bibliographic Services, as well as the software packes \textsc{numpy} \citep{numpy} and \textsc{matplotlib} \citep{matplotlib}.
\section*{Data Availability}
The \textit{NICER} and \textit{XMM-Newton }observations are publicly available on HEASARC, and the \textit{Swift}/BAT \citep{Krimm13} lightcurves can be downloaded from \url{https://swift.gsfc.nasa.gov/results/transients/SMCX-1/}




\bibliographystyle{mnras}
\bibliography{smcx1.bib} 





\bsp	
\label{lastpage}
\end{document}